\documentclass[twocolumn,showpacs,amsmath,amssymb,nofootinbib,floatfix]{revtex4}
\usepackage{graphicx}% Include figure files
\usepackage{dcolumn}%
\usepackage{bm}% bold math
\usepackage{setspace}
\begin{document}
\title{Calculation of complex DNA damage induced by ions}
\author{Eugene Surdutovich$^{1,2}$, David C. Gallagher$^1$, and Andrey V. Solov'yov$^2$\footnote{On leave from A.F. Ioffe Physical Technical Institute, St. Petersburg,
Russia}} \affiliation{ $^1$Department
of Physics, Oakland University, Rochester, Michigan 48309, USA\\
$^2$Frankfurt Institute for Advanced Studies, Ruth-Moufang-Str. 1,
60438 Frankfurt am Main, Germany }
\date{\today}
\begin{abstract}
This paper is devoted to the analysis of the complex damage of DNA
irradiated by ions. The analysis and assessment of complex damage is
important because cells in which it occurs are less likely to
survive because the DNA repair mechanisms may not be sufficiently
effective. We studied the flux of secondary electrons through the
surface of nucleosomes and calculated the radial dose and the
 distribution of clustered
damage around the ion's track. The calculated radial dose
distribution is compared to simulations. The radial distribution of
the complex damage is found to be different from that of the dose.
Comparison with experiments may solve the question of what is more
lethal for the cell, damage complexity or absorbed energy. We
suggest a way to calculate the probability of cell death based on
the complexity of the damage. This work is done within the framework
of the phenomenon-based multiscale approach to radiation damage by
ions.
\end{abstract}


\maketitle

\section{Introduction: multiscale approach to radiation damage}
\label{sec:1}

Ion beam cancer therapy has been in a stage of booming development
recently. Despite the success of this technique, a number of
scientific questions on the microscopic level have not yet been
resolved. This field has attracted much attention in the scientific
community~\cite{Kraft07,Hiroshiko,FokasKraft09,SchardtRMP10,Durante10,Radam09editorial,pre,mutat,SYS}.
Among these is the multiscale approach to the radiation damage
induced by irradiation with ions, aimed at the phenomenon-based
quantitative understanding of the scenario from the incidence of an
energetic ion on tissue to the cell death. This approach joins
together many spatial, temporal, and energetic scales involved in
this scenario. The success of this approach will provide a
phenomenon-based foundation for ion-beam cancer therapy, radiation
protection in space, and other applications of ion beams. The main
issues addressed by the multiscale approach are ion stopping in the
medium~\cite{epjd}, production and transport of secondary electrons
produced as a result of ionization and excitation of the
medium~\cite{epjd,Scif}, interaction of secondary particles with
biological molecules, most important being DNA~\cite{pre}, the
analysis of induced damage, and evaluation of the probabilities of
subsequent cell survival or death. This approach is
interdisciplinary, since it is based on physics, chemistry, and
biology. Moreover, it spans several areas within each of these
disciplines.

The multiscale approach started with the analysis of ion
propagation, which resulted in the
  description of the Bragg peak and the energy spectrum of secondary
  electrons~\cite{epjd,Scif}. The practical goal of these
  works provided a recipe for an economical calculation of the Bragg
  peak position and shape. Theoretically, they concluded that the
   cross section of ionization of molecules of the medium,
   singly-differentiated with respect to the energies of secondary
   electrons, is the most important physical input on this scale
   (the longest in distance and highest in energy). Relativistic
   effects play an important role in describing the position of the
   Bragg peak as well as the excitation channel in
   inelastic interactions~\cite{epjd}. The effect of charge transfer and
   projectile scattering influence the shape of the Bragg peak~\cite{epjd}.
   The effects of
   nuclear fragmentation happening in the events of projectile
   collisions with the nuclei of the medium are also important on this
   scale.

The next scale in energy and space is related to the transport of
the secondary particles, which has been
  considered in Refs.~\cite{pre,EmaRadam09}, but it may still be
  revisited. The results of these analyses give the spatial
  distributions of secondary particles as well as an accurate radial
  dose distribution.

The goal of the analysis of DNA damage mechanisms
  is to obtain the effective cross sections for the dominant
  processes, which should be taken into account in order to calculate
  the probability of different lesions caused by different
  effects.
The above three stages of processes, represent not only different
spatial scales, but also different time scales, ranging from the
$10^{-21}$ to $10^{-5}$ seconds. The aim of the physical part of the
analysis is the calculation of the spatial distribution of primary
DNA damage, including the
  degree of complexity of this damage.
Then, the repair and other biological effects
  can be included and thus the relative biological effectiveness (RBE)
  can be calculated.
The RBE~\cite{Kraft07,Durante10} is one of the key integral characteristics of the effect of
ions compared to that of photons. This ratio compares the doses of
different projectiles leading to the same biological effect.
%The
%calculation of RBE using the multiscale approach
%  will be a result of a constructive quantitative analysis
%  to physical, chemical, and biological phenomena and its predictive
%  power will be on solid theoretical ground. The conditions
%  related to radiation damage may vary, if, e.g., the dose
%  deposition is fast, as in laser-driven beams, chemically active
%  components increasing the number of active agents are present, or
%  biological factors are more important, etc. The multiscale approach,
%  capable of including these variations, will be more versatile than
%  the existing approaches to calculating the RBE.

Traditionally, the radial dose, calculated in
Ref.~\cite{EmaRadam09}, is related to the radial distribution of
damage. However, this does not include the complexity of damage,
which may not be directly related to the dose. It is still not clear
how to relate the dose with the complexity of the damage. This work
is a step in this direction.

Finally, the analysis of the possibility of thermo-mechanical damage
pathways has been started in Refs.~\cite{epjd} and has further
advanced in Refs.~\cite{preheat,prehydro}. This idea stems from
the fact that the energy lost by an ion is transferred to the
tissue's internal degrees of freedom and then becomes thermalized.
We analyzed this transition in Ref.~\cite{preheat} and used it as an
initial condition for hydrodynamic expansion described by a
cylindrical shock wave in Ref.~\cite{prehydro}. These works predict
a rapid rise of temperature and pressure in the vicinity of the
track. Then, when the expansion starts, the pressure is high on the
wave front, but quickly drops in the wake of the wave causing large
pressure gradients, and therefore, strong forces, which may rupture
bonds of biomolecules that may be located within several nm of the
track. It was shown that these forces can be strong enough to break
covalent bonds (more than 10~nN) but act only for a very short time.
An estimate of work done by this force, based on
Ref.~\cite{prehydro}, is several eV, but still more research is
needed in order to investigate whether this represents a separate
mechanism of damage. This effect may also be important in defining
the conditions of the medium in which the other known radiation
damage mechanisms (e.g., electron attachment or free radical attack
on DNA) take place.

This work is devoted to the calculation of damage complexity and
its distribution. This is an
important stage in the multiscale approach, since it is closely
related to the probability of cell death as a result of
damage~\cite{Lynn1,Lynn2,Lynn11,Ward1,Ward2}. Damage complexity is
 one of the defining factors in calculating RBE.

In Sec.~\ref{sec:2} we define the complex damage and present a way
to quantify it. In Sec.~\ref{subsec:2.1} we calculate the fluence of
secondary electrons as a step in the assessment of complex damage.
In Sec.~\ref{subsec:2.2} we calculate the radial dose distribution
and give an example of a calculation of the complex damage on that
basis.

\section{Distributions of the complex damage}
\label{sec:2}

Complex damage is defined as the number of DNA lesions, such as
double strand breaks (DSB), single strand breaks, abasic sites,
damaged bases, etc., that occur within about two helical turns of a
DNA molecule so that, when repair mechanisms are engaged, they treat
a cluster of several of these lesions as a single damage
site~\cite{Lynn1,Lynn2,Lynn11}. In Ref.~\cite{SYS}, the complexity
of DNA damage has been quantified by defining a cluster of damage as
a damaged portion of a DNA molecule by several independent agents,
such as secondary electrons, holes, or radicals.

 In humans, DNA molecules are by and large located in cell
nuclei, where they are organized with proteins into chromatin fibres. The
main structural unit of chromatin fibres is a nucleosome~\cite{nucleosome}.
A nucleosome
core particle consists of about a 146-bp section of a DNA molecule
wrapped around a cylindrical aggregate of eight histone
 proteins (histone octamer).
 %Despite the complexity of its
 %shape~\cite{nucleosome}, for the estimates done in this work, we
 %will assume that the octamer has
 %a cylindrical shape of radius 3.5~nm and height of 6~nm. This
%octamer is wrapped by two loops of a DNA molecule, making the total
%radius of the cylinder to be 5.75~nm.

\subsection{Damage complexity distribution from the random walk approach}
\label{subsec:2.1}

In Ref.~\cite{pre}, we studied the transport of secondary electrons to
a given DNA convolution.
This study led to the calculation of the radial distribution of
DSBs with respect to the ion track. This calculation was limited by
only considering secondary electrons to be the agents of DNA
lesions. Nevertheless, this allowed us to make an estimation of the
number of DSBs produced by ions per unit length of track in the
vicinity of the Bragg peak. The results obtained in that work were
in reasonable agreement with the experimental data~\cite{Jakob}. The
approach of Ref.~\cite{pre} can be used for calculating the radial
distribution of damage complexity.

Let us choose two adjacent convolutions of a DNA molecule as a target.
 Then,  the average number
of lesions per this segment of DNA, $N$, is given by a product of the
probability of inducing damage by a secondary particle on impact, $\Gamma$,
 by the
fluence through the target. Alternatively, it is given by the same
probability multiplied by the volume of the segment and by the number
density of agents. The probability of complex damage is then
a Poisson distribution $P(N,\nu)$,
\begin{eqnarray}
P(\rho, \nu)=\exp\left({-N(\rho)}\right) \frac{N(\rho)^\nu}{\nu !}~,
\label{distrad1}
\end{eqnarray}
where $\nu$ is the degree of
complexity~\cite{SYS}. In Eq.(\ref{distrad1}), $N$ is written as a
function of $\rho$, the distance of the segment from the track. Our
goal is to calculate the radial distribution
of complex damage with respect to the ion track in the simplest
case, when all agents are equivalent, keeping the probability $\Gamma$
as a parameter.
%this distribution is given by
%where $N(\rho)$ is the average number of lesions per DNA segment
%at a given distance $\rho$ from the track.
%Besides the
%volume of the cluster, $N(\rho)$ contains two more parameters: these
%are the number density of agents at $\rho$ and the probability of
%inflicting a lesion. They always appear as a product and therefore
%they can be considered as a single parameter. Still, it is desirable
%to calculate both of them. The number density can be obtained from
In this paper, we limit secondary particles to secondary electrons. A
further development will include transport of secondary particles
including chemical reactions and more details of their
distributions.

In this section, we calculate the fluence of the secondary electrons, through the DNA segment.
In order to do this, we consider their diffusion
from the place of their origin as was done in Ref.~\cite{pre}.
We assume that the diffusion of the secondary
electrons is cylindrically symmetric with respect to the ion's track
and calculate the number of electrons, which hit two adjacent
convolutions of a DNA molecule.
%the fraction of the surface of the
%cylinder facing the ion's track corresponding to two adjacent
%convolutions of a DNA molecule.
The cylindrical diffusion in the
vicinity of the Bragg peak can be justified by the fact that during
the time that it takes secondary electrons to diffuse by about
10~nm, the projectile moves a distance of about
1~$\mu$m~\cite{prehydro}. The linear energy transfer (LET) along
this distance, described by the coordinate $\zeta$, remains nearly
constant, as well as the production of secondary particles per unit
length $\frac{dN}{d\zeta}$; therefore, the latter is independent of
${\zeta}$. This number of secondary electrons produced per one nm of
the ion's track is taken to be equal to 20, which corresponds to the
average number of ionizations per one nm of ion's track in the
vicinity of the Bragg peak~\cite{pre,epjd}.

\begin{figure}
\includegraphics[width=3.5in]{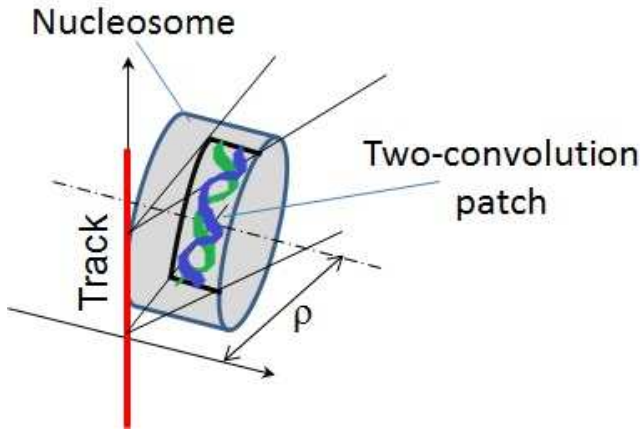}
\caption{Geometry of the problem. Secondary electrons radially
diffuse from the ion's track and interact with a section of DNA
molecule wrapped around a histone octamer.}
\label{fig.umbra}       % Give a unique label
\end{figure}

Naturally, we expect the largest damage occurring when the incident
ion passes through a nucleosome. Therefore, in this paper, we
calculate the complex damage that takes place in two
consecutive convolutions of a DNA molecule
on the surface of a nucleosome situated outside the ion's track (neglecting
the stretches of linker DNA connecting nucleosomes).
In what follows, a nucleosome is represented by a cylinder of
 radius 5.75~nm and height of 6~nm and the target section of a
 DNA molecule is a rectangular
 patch ($7.2{\rm nm}\times2.3{\rm nm}$) of
its surface, as shown in
Fig.~\ref{fig.umbra}.

In order to calculate the fluence, we consider the
 rate of secondary electrons, $dN_A/dt$, at
the time $t$, passing through the patch $d{\vec A}$, located at a
distance $\rho$ from the track. According to
Ref.~\cite{Chandra}, for a cylindrically symmetric random walk, it
is given by the expression
\begin{eqnarray}
\frac{dN_A(\vec r, t)}{dt} = d{\vec A}\cdot D \nabla P(t,
\rho)\frac{d N}{d\zeta}\nonumber\\= d{\vec A}\cdot D { \bf
n_\rho}\frac{\partial P(t, \rho)}{\partial
  \rho}\frac{d
N}{d\zeta}~, \label{mult1t}
\end{eqnarray}
where $D={\bar v} l/4$ is the diffusion coefficient,
$l$ is the elastic mean free path of electrons in the medium\footnote{In
two dimensions, $l$ is a product of the mean free path in three dimensions
multiplied by the factor of $\sqrt{2/3}$.}, ${\bar v}$ is
the speed of the electron, $\bf n_\rho$ is a unit vector in the radial
direction from the track, and
\begin{eqnarray}
P(t, \rho)=\frac{1}{\pi
    {\bar v}t l}\exp\left(-\frac{\rho^2}{{\bar v}t l}\right)
\label{rwalk2t}
\end{eqnarray}
is the probability density to observe a randomly walking electron at
a time $t$ and a distance $\rho$ from the track. Eventually we are
going to integrate Eq.~(\ref{mult1t}) over both, the time (to get
the total number of electrons incident on the patch $d{\vec A}$) and
$d{\vec A}$ (in order to calculate the total number of electrons
incident on a two-twist-segment of a DNA molecule). Before we do
this, we need to somewhat modify expressions (\ref{mult1t}) and
(\ref{rwalk2t}).

First, the time dependence can be translated to the dependence on
number of steps, $k$, using ${\bar v}t=kl$ and ${\bar v}dt=l dk$.

Second, there is a probability that the electron interacts with a
molecule inelastically, loses energy and drops off from a random
walk. In order to account for such a subtraction, we introduce an
attenuation factor $\epsilon(k)$. In Ref.~\cite{pre}, we used
\begin{eqnarray}
\epsilon(k)=\gamma \exp(-\gamma k)~, \label{epsilon1}
\end{eqnarray}
where $\gamma$ is a constant, proportional to the ratio of mean free
paths between inelastic and elastic collisions. This expression is
physically motivated, but it does not take into account the energy
dependence of mean free paths and their ratio. In this paper, we
will keep the elastic mean free path $l$ energy-independent and
equal to 1~nm~\cite{EmaRadam09}, while we will use the
attenuation given by
\begin{eqnarray}
\epsilon(k)=\exp{\left(\alpha \exp{\left(-k^\beta\right)}\right)}~.
\label{epsilon2}
\end{eqnarray}
This expression with constants $\alpha=60$ and $\beta=0.055$ appears
as a result of fitting the radial dose distribution derived from a
model of secondary electron transport to that obtained using Monte
Carlo simulations~\cite{Plante}. This model assumes a random walk of electrons
with a constant mean free path, i.e., the same used by us in this
section.
 Expression~(\ref{epsilon2}), with a modified dependence on $k$,
implicitly introduces the dependence of the attenuation on energy.
The attenuation according Eq.~(\ref{epsilon2}) is steeper than that
according to Eq.~(\ref{epsilon1}) for small $k$. This means that
electrons with higher energy tend to lose it in inelastic collisions
more quickly than those with smaller energies and the attenuation at
large $k$ is much smaller.
 We will return to this parametrization in the following
section.

\begin{figure}
\includegraphics[width=3.5in]{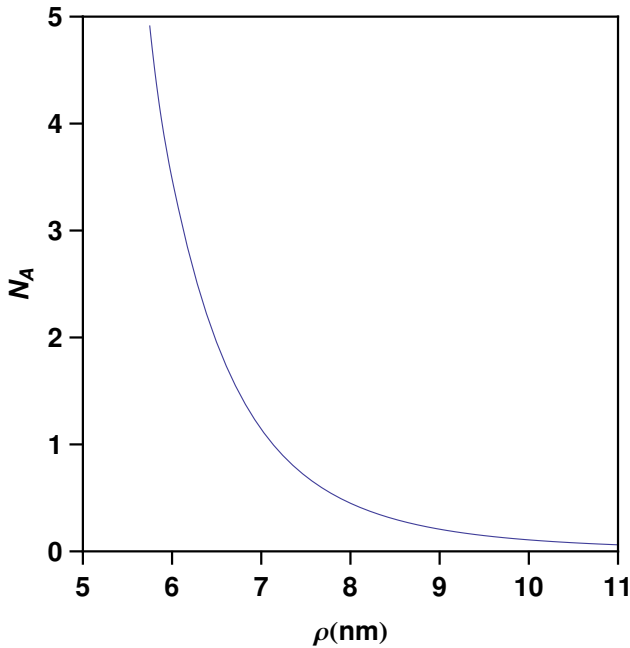}
\caption{Number of secondary electrons diffused through a
two-convolution segment of DNA molecule on the surface of a
nucleosome in the
  vicinity of the ion's track, plotted as a function of the distance
  $\rho$ of nucleosome from the track.
  The calculation is done with a use of the attenuation
  function given by Eq.~(\ref{epsilon2}).}
\label{fig.nrad}       % Give a unique label
\end{figure}

 Now we can rewrite Eq.~(\ref{mult1t}), substituting
(\ref{rwalk2t}), including the attenuation,  and switching from
variable $t$ to $k$ as
\begin{eqnarray}
dN_A(\vec r, k)= dk d{\vec A}\cdot { \bf n_\rho}
    \frac{\rho}{2\pi k^2 l^2} \exp\left(-\frac{\rho^2}{k l^2}\right)\epsilon(k)\frac{d
N}{d\zeta} \label{mult4}
\end{eqnarray}
and integrate it over the target part of surface of the cylinder,
representing a nucleosome.
The results of integration~(\ref{mult4}) over time and the area of
the patch are shown in Fig.~\ref{fig.nrad}. As expected, this number
decreases with distance $\rho$ from the track.

If we multiply this number, $N_A$, by
the probability, $\Gamma$, of producing a lesion in a DNA molecule,
uniformly distributed on the surface of the nucleosome, we obtain
the dependence of the number of lesions on the distance from the
track. Then, Eq.~(\ref{distrad1}) can be used, with
\begin{eqnarray}
N(\rho)=\Gamma N_{A}(\rho) \label{mult4plus}
\end{eqnarray}
 to calculate the radial distributions of
probabilities of clusters of lesions.

This number has to be
corrected by to include further ionizations and
holes, which also play a role in the damage. Holes may recombine
producing Auger electrons, capable of inducing lesions to
DNA~\cite{prauger}. Before these corrections are made,
these calculations remain qualitative.
% because of the unknown
%parameter $\xi$.
When the transport properties of electrons and
other secondary particles in the medium are known and $N(\rho)$ is
calculated more definitely, the approach to the calculation of the
clustered damage, described above, can be useful.

In addition to this calculation, it is possible to compute similar
probabilities for cases when the track passes through the
nucleosome; but this would require calculating the transport of
secondary electrons through a histone and knowledge of the elastic
and inelastic cross sections of electrons in this medium. Recent
calculations indicate a 20\% higher stopping power of DNA compared
to liquid water~\cite{Molina}. We will postpone these calculations
until another time. However, it may be worth mentioning that the
clustered damage of a histone may also deserve attention in regard
to cell damage.

%The results, shown in Fig.~\ref{fig.nrad},
%  support the choice of a
%nucleosome of diameter 6~nm as a host for clustered damage, since
%the number of secondary electrons, incident on a two-convolution
%segment, drops by the factor of 10 when the distance of a nucleosome
%from the track increases by 3~nm. Also, the characteristic ranges of
%propagation of secondary electrons with energies below 50~eV are
%about 10~nm~\cite{Meesungnoen02}. Finally, the clustered damage is
%biologically significant if the lesions forming a cluster are close
%enough together, e.g., on the distance (along the DNA) of less than
%about 20~bp~\cite{Lynn1,Lynn2,Lynn11}. The length of such a segment
%of a DNA molecule is about 7~nm and this is of the order of the
%linear size of the face of the nucleosome exposed to radiation.
%++++++++++++++++++++++++++++++++++++++++++++++++++++

The calculation of the number of secondary electrons passing through
a patch on a nucleosome presented in this section is important for
several reasons. First, it can be compared with Monte Carlo
simulations done for the purposes of nanodosimetry~\cite{Marion}.
Second, it will be possible to compare this dependence
(correspondingly modified) with dosimetric
experiments~\cite{Roz1,Roz2}. At this point, it is possible to use
the dependence shown in Fig.~\ref{fig.nrad} for the calculation of
the complex damage (using additional parameters); however, in this
paper, we choose to use the radial dose distribution to demonstrate
a calculation of complex damage. At this moment, the latter approach
allows for more checkpoints and we describe it in the next section.

\subsection{Derivation of damage complexity from the radial dose distribution}
\label{subsec:2.2}

As was shown in Ref.~\cite{EmaRadam09}, the radial number density
distribution of secondary electrons that lost energy and became
thermalized or bound is related to the radial dose. Here, we revisit
the calculation of the radial dose and infer the secondary particle
distribution, with the complexity distribution following from that.

\begin{figure}
\includegraphics[width=3.5in]{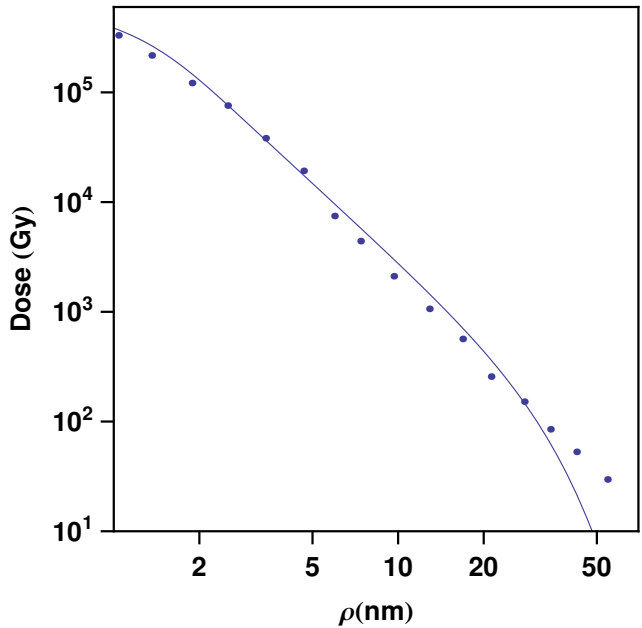}
\caption{Comparison of the calculated radial dose (line),
Eq.~(\ref{radDose})
 with the simulated in Ref.~\cite{Plante} (dots) for 25 MeV/u carbon ions.}
\label{fig.plante}       % Give a unique label
\end{figure}

Let us assume that all secondary electrons start from the ion's
track and propagate via random walk in two dimensions; this
corresponds to the cylindrically symmetric propagation (neglecting
some fast $\delta$-electrons). Then, according to
Eq.~(\ref{rwalk2t}), rewritten in terms of $k$, the probability to
find a secondary particle in
a cylindrical layer of unit length between $\rho$ and $\rho+d\rho$
after $k$ random steps is $\frac{dN_s}{d\zeta}P(k,\rho)2\pi \rho
d\rho $. This probability is normalized to $\frac{dN_s}{d\zeta}$,
for any number of steps $k$ if we integrate over $d\rho$ from 0 to
infinity. The normalization does not change if we include the
attenuation $\epsilon(k)$ due to inelastic processes and introduce a
distribution over $k$. Transferring from the sum to the integral,
appropriate for large $k$, this can be written as
\begin{eqnarray}
\int_1^\infty \int_0^\infty P(k,
\rho)\epsilon(k) 2\pi \rho d\rho dk=1~.
\label{rwalk2norm}
\end{eqnarray}
Then the density of the particles, which lost energy within the
cylindrical layer of a unit length between $\rho$ and $\rho+d\rho$
can be obtained by dividing the integrand of Eq.~(\ref{rwalk2norm})
by the volume of this shell of a unit length, i.e., $2\pi \rho
d\rho$ and the radial dose ${\cal D}(\rho)$ can be obtained by
multiplication of this density by the average energy per particle
${\bar W}=45$~eV~\cite{epjd}:
\begin{eqnarray}
{\cal D}(\rho)={\bar W}\frac{dN_s}{d\zeta}\int_1^\infty P(k, \rho)
\epsilon(k)dk~. \label{radDose}
\end{eqnarray}
This dose is normalized by the LET:
\begin{eqnarray}
\int_0^\infty {\cal D}(\rho)2\pi \rho d\rho=LET~. \label{radDoseNorm}
\end{eqnarray}
However, the radial dose distribution calculated using
Eq.~(\ref{radDose}) with the attenuation, defined by
Eq.~(\ref{epsilon1}),
 does not agree with simulations of, e.g.,
Ref.~\cite{Plante}. The reason for this is that in
Eq.~(\ref{epsilon1}) we have assumed energy-independent attenuation.
According to, e.g., Ref.~\cite{Tung}, both elastic and inelastic
mean free paths are energy dependent. Moreover, as has been pointed
out in Ref.~\cite{pre}, the dependence of ranges of low-energy
electrons in liquid water on energy, discussed in
Ref.~\cite{Meesungnoen02}, indicates that the attenuation steeply
decreases as the energy of the electron decreases (after several
inelastic collisions). This can be taken into account by
parametrizing the attenuation as a function of $k$ so that the dose
calculated using Eq.~(\ref{radDose}) agrees with experiments and
simulations. In this work, we have found that the dose, calculated
using Eq.~(\ref{radDose}) with attenuation defined by
(\ref{epsilon2}) with parameters $\alpha$ and $\beta$ given
above\footnote{Expression (\ref{epsilon2}) has to be divided by
$\int \epsilon(k)dk $ for normalization.} is in reasonable agreement
with that simulated in Ref.~\cite{Plante}. This comparison is shown
in Fig.~\ref{fig.plante}.

\begin{figure}
\includegraphics[width=3.5in]{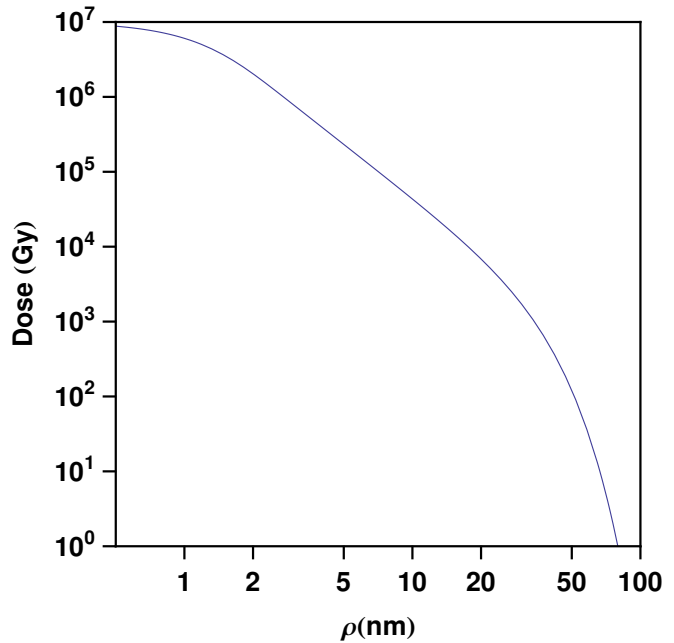}
\caption{Calculated radial dose distribution for LET=0.9~keV/nm.}
\label{fig.dose}       % Give a unique label
\end{figure}

The simulation, done in Ref.~\cite{Plante} corresponds to 25-MeV
carbon ions with $LET=60$~eV/nm. This is about 4~mm away from the
Bragg peak, where $LET=900$~eV/nm. Therefore we recalculated the
same dose using the procedure described above for 0.3-MeV/u carbon
ions with a $LET=900$~eV/nm. The result is presented in
Fig.~\ref{fig.dose}.
This distribution can be compared to the experimental measurements
of the radial dose distribution; however, at this moment, such data
are not available for such a LET.

Using this dose distribution around a single ion's track, we can
calculate the distribution of clusters of DNA damage. In order to do
this, we have to divide the expression~(\ref{radDose}) for the dose
by ${\bar W}$, and obtain the radial distribution of the density of
inelastically interacting secondary electrons. If then we multiply
that by the effective volume of the target segment DNA and the
probability of producing a lesion $\Gamma$, we will obtain
$N(\rho)$. Then, we can calculate the radial distribution of complex
damage using Eq.~(\ref{distrad1}). This will only be correct if
$N(\rho)$ does not significantly change over this volume.

\begin{figure}
\includegraphics[width=3.5in]{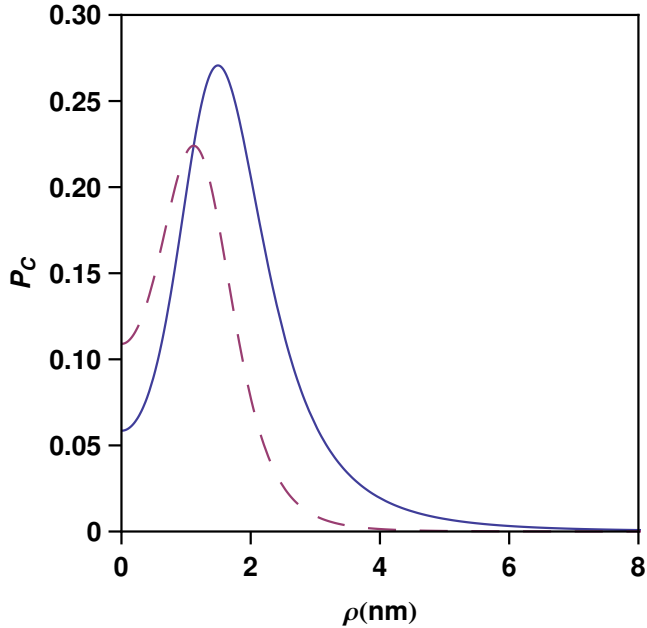}
\caption{Radial distribution of clusters of two (solid line) and
  clusters of three lesions (dashed line) for $\Gamma=0.1$.}
\label{fig.comp2}       % Give a unique label
\end{figure}

If we assume that the effects of varying $N(\rho)$ on the size of
some effective volume can be neglected, then we can calculate the
radial distribution $P_c$ of  clusters for a given $\nu$. An example
of such dependencies for a volume of 40~nm$^3$ (corresponding to the
volume occupied by two convolutions of DNA molecule) and
$\Gamma=0.1$ of two- and
 three-lesion clusters, are shown in Fig.~\ref{fig.comp2}.

\begin{figure}
\includegraphics[width=3.5in]{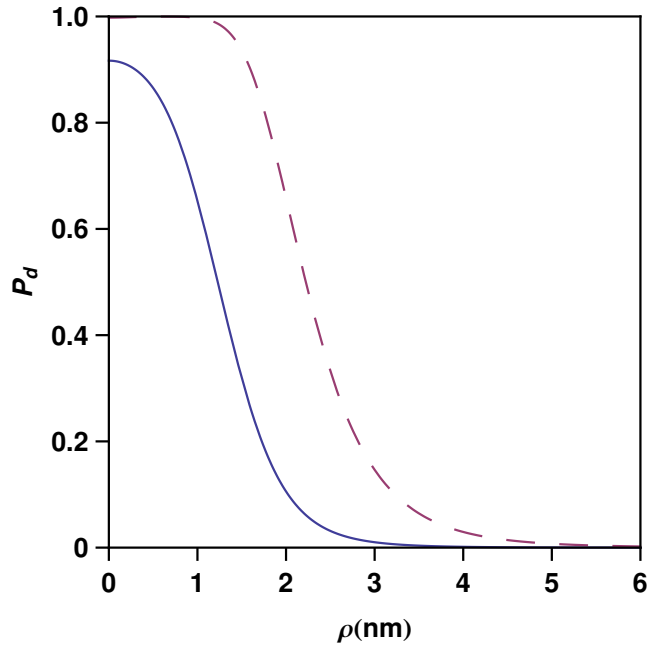}
\caption{Radial distribution of clusters of three and more lesions,
  deemed proportional to the probability of cell death. The solid curve
  corresponds to $\Gamma=0.1$ and the dashed one to $\Gamma=0.3$.
 }
\label{fig.death}       % Give a unique label
\end{figure}

%+++++++++++++++++++++++++++++++++++++++++++++++++++++++++
These distributions give us an opportunity to verify the
significance of clusterization. If, e.g., all clusters containing
three and higher lesions are lethal for the cell, we can add up
their probabilities and plot the dependence of the probability of
cell death, $P_d$, on the distance from the track. These
dependencies for $\Gamma=0.1$ and $\Gamma=0.3$ are shown in
Fig.~\ref{fig.death}.
This figure indicates that if the clusters of three and more lesions
per nucleosome are indeed lethal, then the effective distance from
the track on which the cells are killed is less than 1.5~nm for
$\Gamma=0.1$ and it exceeds 2~nm for $\Gamma=0.3$. Hence, in the
first case, it is essential that the ion passes through a nucleosome
in order to kill the cell, while in the second, a nucleosome can be
at a distance and still be severely damaged. This analysis opens
several fields for comparison with experiments: the dependence of
lethality on the radial distance from the track and on the size of
clusters of lesions for biophysics and the the radial dependence of
dose and cluster damage distribution for nano-dosimetry. The radial
scale in Fig.~\ref{fig.death} is shorter than 10~nm. Even though,
this size is about 1000 times smaller (for glial cells) than that of
the cell's nucleus~\cite{pre}; it plays a significant role in
calculations of the probability of cell death and will be critical
for the comparisons with nano-dosimetric
data~\cite{Roz1,Roz2,dosim1,Marion}.

\begin{figure}
\includegraphics[width=3.5in]{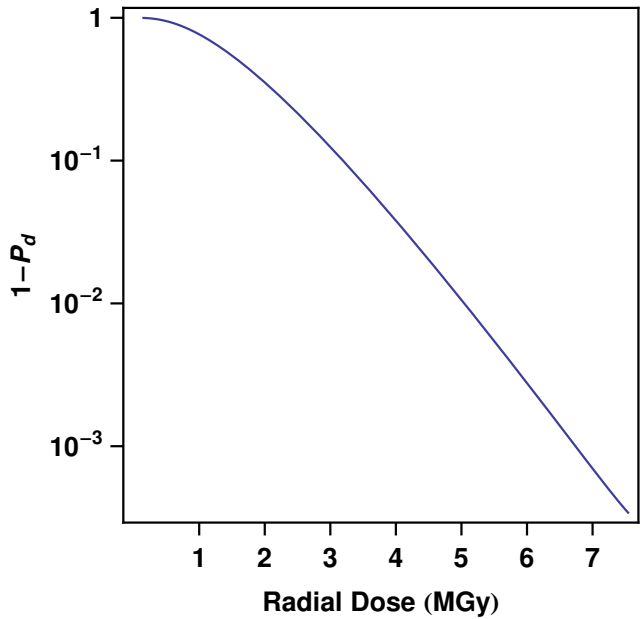}
\caption{The dependence of $1-P_d$, similar to cell survival rate
(dimensionless)
 on the local radial dose per ion with $\Gamma=0.3$.
 }
\label{fig.surv}       % Give a unique label
\end{figure}

If we keep the assumption that three and higher order lesion
clusters are lethal to the cell, then we can plot the dependence of
$1-P_d$, similar to the probability of cell survival, on the radial
dose. This dependence is presented in Fig.~\ref{fig.surv} and this
can also be compared with experiments.
The scale on the abscissa for the radial dose is indeed in MGy. This
is by a factor of $10^6$ larger than that in typical cell survival
curves~\cite{Kraemer1}, where the dose is absolute, i.e.,
planar-integrated per ion and distributed for the whole beam.  A
typical spatial distance between the ion tracks is larger than
350~nm.\footnote{According to the beam data~\cite{Kraft07}.}
Therefore, the volume per one nm of the ion's track is at
least $10^5$~nm$^3$. This makes the average integral dose at the
Bragg peak of a single carbon ion about $7\times10^{-3}$~eV
nm$^{-3}=110$~Gy. This dose has to be averaged once again when
one considers the whole ion beam. This brings another factor of the
order of 0.1, which reduces the maximum dose to 10-15~Gy. The
absolute dose is important for treatment planning, but here we are
interested in a more detailed description and the radial, not
averaged dose, is more relevant for this purpose. This is why we
%The
%average effective volume per ion is many times larger than that of a
%cylinder, which we considered in this work, and this makes the
%absolute dose many times smaller.
%Such an integration requires
%further assumptions and would make results less general. Therefore,
%we
presented the dependence of a probability cell survival on the local
radial dose. Below we show how to do a planar integration and give
an example of an estimate.

The radial distributions of clustered damage probabilities can be
integrated over the radius in order to obtain the probability of
lethal damage per unit length.
%Still another path can be taken if we decide to ignore the radial
%distribution of dose or number density and just consider the
%longitudinal distribution of clusters.
This is relevant for current experiments. When experimentalists
study foci, which reveal the efforts of proteins to fix damaged DNA,
they observe that the foci are very large, compared to the scale of
the radial distribution of the dose. The experimentalists can
measure the linear density of clusters along the track and
hypothesize about the number of certain lesions, such as DSB, per
unit length~\cite{Jakob1}.

In order to obtain the longitudinal distributions of clusters, we
have to introduce the density of the distribution of nucleosomes
with respect to the ion track, $\eta(\rho)$. Then, we can integrate
the radial-dependent probability of the complex damage given by
Eq.~(\ref{distrad1}) (for appropriate $N(\rho)$ dependence) with
this density distribution:
\begin{eqnarray}
P(\nu)=\int_0^\infty \exp{\left(-N(\rho)\right)}
\frac{N(\rho)^\nu}{\nu !}\eta(\rho) 2\pi \rho d\rho~.
\label{distrad-long}
\end{eqnarray}
This gives the number of clusters of $\nu$ lesions per nm, which can
be compared with the nano-dosimetric
experiments~\cite{Roz1,Roz2,dosim1,Marion} and can give still
another relation for unknown parameters such as $\Gamma$ and the
dependence of lethality of damage on the order of cluster $\nu$.

The density of the distribution of nucleosomes, $\eta(\rho)$,
depends on the structure of packing nucleosomes in fibers. If we
consider a  section of a cylindrical fiber of tightly packed
nucleosomes~\cite{nucleosome} to be parallel to the track 1~nm away
from its surface, then an estimate made with the above assumptions
producing the maximal effect of damage complexity, predicts about 3
complex damage sites (with $\nu > 2$) per
 10~nm of a carbon ion's track.

\section{Conclusions}
\label{sec:3}

The multiscale approach was designed in order to understand the
mechanisms that make the ion-beam therapy effective. This includes
the understanding of what is truly different between different
therapies. It is widely accepted that the high-LET radiation brings
about high dose in the desired location. However, it is not yet
clear whether the dose entirely accounts for all biological
consequences. Namely, how different are the dose and complexity
distributions and which of them is responsible for the cell death?
This paper tackled these questions and, although more experiments
are needed to confirm them, the principle framework of the problem
has been set up.

The main accomplishments of this paper are the calculation of the
radial dose distribution (comparable with that obtained by
simulations), derivation of the radial distribution of secondary
electrons from the radial dose distribution, and the calculation of
the radial distributions of different clusters of lesions. On the
basis of these distributions we developed models for calculations of
dependencies of the probabilities of cell death as a result of
complex damage of DNA on the distance from the ion's track and along
the track. These calculations may be very practical and we hope that
it will be explored by experimentalists in nano-dosimetry as well as
by biophysicists. The main principle point in our approach to damage
complexity is that it can be described by a spatial distribution and
compared  to the radial dose distribution and the distribution of
killed cells.

\acknowledgments{ We are grateful to the support of the authors'
collaboration by the Deutsche Forschungsgemeinschaft. ES thanks J.S.
Payson for a constructive critique. }

%\bibliography{bibliography1}

\begin{thebibliography}{33}
\expandafter\ifx\csname natexlab\endcsname\relax\def\natexlab#1{#1}\fi
\expandafter\ifx\csname bibnamefont\endcsname\relax
  \def\bibnamefont#1{#1}\fi
\expandafter\ifx\csname bibfnamefont\endcsname\relax
  \def\bibfnamefont#1{#1}\fi
\expandafter\ifx\csname citenamefont\endcsname\relax
  \def\citenamefont#1{#1}\fi
\expandafter\ifx\csname url\endcsname\relax
  \def\url#1{\texttt{#1}}\fi
\expandafter\ifx\csname urlprefix\endcsname\relax\def\urlprefix{URL }\fi
\providecommand{\bibinfo}[2]{#2}
\providecommand{\eprint}[2][]{\url{#2}}

\bibitem[{\citenamefont{Amaldi and Kraft}(2007)}]{Kraft07}
\bibinfo{author}{\bibfnamefont{U.}~\bibnamefont{Amaldi}} \bibnamefont{and}
  \bibinfo{author}{\bibfnamefont{G.}~\bibnamefont{Kraft}}, \bibinfo{journal}{J.
  Radiat. Res.} \textbf{\bibinfo{volume}{48}}, \bibinfo{pages}{A27}
  (\bibinfo{year}{2007}).

\bibitem[{\citenamefont{Tsujii et~al.}(2008)\citenamefont{Tsujii, Kamada, Baba,
  Tsuji, Kato, Kato, Yamada, Yasuda, Yanagi, Kato et~al.}}]{Hiroshiko}
\bibinfo{author}{\bibfnamefont{H.}~\bibnamefont{Tsujii}},
  \bibinfo{author}{\bibfnamefont{T.}~\bibnamefont{Kamada}},
  \bibinfo{author}{\bibfnamefont{M.}~\bibnamefont{Baba}},
  \bibinfo{author}{\bibfnamefont{H.}~\bibnamefont{Tsuji}},
  \bibinfo{author}{\bibfnamefont{H.}~\bibnamefont{Kato}},
  \bibinfo{author}{\bibfnamefont{S.}~\bibnamefont{Kato}},
  \bibinfo{author}{\bibfnamefont{S.}~\bibnamefont{Yamada}},
  \bibinfo{author}{\bibfnamefont{S.}~\bibnamefont{Yasuda}},
  \bibinfo{author}{\bibfnamefont{T.}~\bibnamefont{Yanagi}},
  \bibinfo{author}{\bibfnamefont{H.}~\bibnamefont{Kato}}, \bibnamefont{et~al.},
  \bibinfo{journal}{New J. Phys.} \textbf{\bibinfo{volume}{10}},
  \bibinfo{pages}{075009} (\bibinfo{year}{2008}).

\bibitem[{\citenamefont{Fokas et~al.}(2009)\citenamefont{Fokas, Kraft, An, and
  Engenhart-Cabillic}}]{FokasKraft09}
\bibinfo{author}{\bibfnamefont{E.}~\bibnamefont{Fokas}},
  \bibinfo{author}{\bibfnamefont{G.}~\bibnamefont{Kraft}},
  \bibinfo{author}{\bibfnamefont{H.}~\bibnamefont{An}}, \bibnamefont{and}
  \bibinfo{author}{\bibfnamefont{R.}~\bibnamefont{Engenhart-Cabillic}},
  \bibinfo{journal}{Biochim. Biophys. Act.} \textbf{\bibinfo{volume}{1796}},
  \bibinfo{pages}{216} (\bibinfo{year}{2009}).

\bibitem[{\citenamefont{Schardt et~al.}(2010)\citenamefont{Schardt,
  Els{\"a}sser, and Schulz-Ertner}}]{SchardtRMP10}
\bibinfo{author}{\bibfnamefont{D.}~\bibnamefont{Schardt}},
  \bibinfo{author}{\bibfnamefont{T.}~\bibnamefont{Els{\"a}sser}},
  \bibnamefont{and}
  \bibinfo{author}{\bibfnamefont{D.}~\bibnamefont{Schulz-Ertner}},
  \bibinfo{journal}{Rev. Mod. Phys.} \textbf{\bibinfo{volume}{82}},
  \bibinfo{pages}{383} (\bibinfo{year}{2010}).

\bibitem[{\citenamefont{Durante and Loeffler}(2010)}]{Durante10}
\bibinfo{author}{\bibfnamefont{M.}~\bibnamefont{Durante}} \bibnamefont{and}
  \bibinfo{author}{\bibfnamefont{J.}~\bibnamefont{Loeffler}},
  \bibinfo{journal}{Nat. Rev. Clin. Oncol.} \textbf{\bibinfo{volume}{7}},
  \bibinfo{pages}{37} (\bibinfo{year}{2010}).

\bibitem[{\citenamefont{Baccarelli et~al.}(2010)\citenamefont{Baccarelli,
  Gianturco, Scifoni, Solov'yov, and Surdutovich}}]{Radam09editorial}
\bibinfo{author}{\bibfnamefont{I.}~\bibnamefont{Baccarelli}},
  \bibinfo{author}{\bibfnamefont{F.}~\bibnamefont{Gianturco}},
  \bibinfo{author}{\bibfnamefont{E.}~\bibnamefont{Scifoni}},
  \bibinfo{author}{\bibfnamefont{A.}~\bibnamefont{Solov'yov}},
  \bibnamefont{and}
  \bibinfo{author}{\bibfnamefont{E.}~\bibnamefont{Surdutovich}},
  \bibinfo{journal}{Eur. Phys. J. D} \textbf{\bibinfo{volume}{60}},
  \bibinfo{pages}{1} (\bibinfo{year}{2010}).

\bibitem[{\citenamefont{Solov'yov et~al.}(2009)\citenamefont{Solov'yov,
  Surdutovich, Scifoni, Mishustin, and Greiner}}]{pre}
\bibinfo{author}{\bibfnamefont{A.}~\bibnamefont{Solov'yov}},
  \bibinfo{author}{\bibfnamefont{E.}~\bibnamefont{Surdutovich}},
  \bibinfo{author}{\bibfnamefont{E.}~\bibnamefont{Scifoni}},
  \bibinfo{author}{\bibfnamefont{I.}~\bibnamefont{Mishustin}},
  \bibnamefont{and} \bibinfo{author}{\bibfnamefont{W.}~\bibnamefont{Greiner}},
  \bibinfo{journal}{Phys. Rev.} \textbf{\bibinfo{volume}{E79}},
  \bibinfo{pages}{011909} (\bibinfo{year}{2009}).

\bibitem[{\citenamefont{Surdutovich
  et~al.}(2010{\natexlab{a}})\citenamefont{Surdutovich, Scifoni, , and
  Solov'yov}}]{mutat}
\bibinfo{author}{\bibfnamefont{E.}~\bibnamefont{Surdutovich}},
  \bibinfo{author}{\bibfnamefont{E.}~\bibnamefont{Scifoni}}, ,
  \bibnamefont{and}
  \bibinfo{author}{\bibfnamefont{A.}~\bibnamefont{Solov'yov}},
  \bibinfo{journal}{Mutat. Res.} \textbf{\bibinfo{volume}{704}},
  \bibinfo{pages}{206} (\bibinfo{year}{2010}{\natexlab{a}}).

\bibitem[{\citenamefont{Surdutovich
  et~al.}(2010{\natexlab{b}})\citenamefont{Surdutovich, Yakubovich, and
  Solov'yov}}]{SYS}
\bibinfo{author}{\bibfnamefont{E.}~\bibnamefont{Surdutovich}},
  \bibinfo{author}{\bibfnamefont{A.}~\bibnamefont{Yakubovich}},
  \bibnamefont{and}
  \bibinfo{author}{\bibfnamefont{A.}~\bibnamefont{Solov'yov}},
  \bibinfo{journal}{Eur. Phys. J. D} \textbf{\bibinfo{volume}{60}},
  \bibinfo{pages}{101} (\bibinfo{year}{2010}{\natexlab{b}}).

\bibitem[{\citenamefont{Surdutovich et~al.}(2009)\citenamefont{Surdutovich,
  Obolensky, Scifoni, Pshenichnov, Mishustin, Solov'yov, and Greiner}}]{epjd}
\bibinfo{author}{\bibfnamefont{E.}~\bibnamefont{Surdutovich}},
  \bibinfo{author}{\bibfnamefont{O.}~\bibnamefont{Obolensky}},
  \bibinfo{author}{\bibfnamefont{E.}~\bibnamefont{Scifoni}},
  \bibinfo{author}{\bibfnamefont{I.}~\bibnamefont{Pshenichnov}},
  \bibinfo{author}{\bibfnamefont{I.}~\bibnamefont{Mishustin}},
  \bibinfo{author}{\bibfnamefont{A.}~\bibnamefont{Solov'yov}},
  \bibnamefont{and} \bibinfo{author}{\bibfnamefont{W.}~\bibnamefont{Greiner}},
  \bibinfo{journal}{Eur. Phys. J. D} \textbf{\bibinfo{volume}{51}},
  \bibinfo{pages}{63} (\bibinfo{year}{2009}).

\bibitem[{\citenamefont{Scifoni
  et~al.}(2010{\natexlab{a}})\citenamefont{Scifoni, Surdutovich, and
  Solovyov}}]{Scif}
\bibinfo{author}{\bibfnamefont{E.}~\bibnamefont{Scifoni}},
  \bibinfo{author}{\bibfnamefont{E.}~\bibnamefont{Surdutovich}},
  \bibnamefont{and} \bibinfo{author}{\bibfnamefont{A.}~\bibnamefont{Solovyov}},
  \bibinfo{journal}{Phys Rev. E} \textbf{\bibinfo{volume}{81}},
  \bibinfo{pages}{021903} (\bibinfo{year}{2010}{\natexlab{a}}).

\bibitem[{\citenamefont{Scifoni
  et~al.}(2010{\natexlab{b}})\citenamefont{Scifoni, Surdutovich, and
  Solov'yov}}]{EmaRadam09}
\bibinfo{author}{\bibfnamefont{E.}~\bibnamefont{Scifoni}},
  \bibinfo{author}{\bibfnamefont{E.}~\bibnamefont{Surdutovich}},
  \bibnamefont{and}
  \bibinfo{author}{\bibfnamefont{A.}~\bibnamefont{Solov'yov}},
  \bibinfo{journal}{Eur. Phys. J. D} \textbf{\bibinfo{volume}{60}},
  \bibinfo{pages}{115} (\bibinfo{year}{2010}{\natexlab{b}}).

\bibitem[{\citenamefont{Toulemonde et~al.}(2009)\citenamefont{Toulemonde,
  Surdutovich, and Solov'yov}}]{preheat}
\bibinfo{author}{\bibfnamefont{M.}~\bibnamefont{Toulemonde}},
  \bibinfo{author}{\bibfnamefont{E.}~\bibnamefont{Surdutovich}},
  \bibnamefont{and}
  \bibinfo{author}{\bibfnamefont{A.}~\bibnamefont{Solov'yov}},
  \bibinfo{journal}{Phys. Rev. E} \textbf{\bibinfo{volume}{80}},
  \bibinfo{pages}{031913} (\bibinfo{year}{2009}).

\bibitem[{\citenamefont{Surdutovich and Solov'yov}(2010)}]{prehydro}
\bibinfo{author}{\bibfnamefont{E.}~\bibnamefont{Surdutovich}} \bibnamefont{and}
  \bibinfo{author}{\bibfnamefont{A.}~\bibnamefont{Solov'yov}},
  \bibinfo{journal}{Phys. Rev. E} \textbf{\bibinfo{volume}{82}},
  \bibinfo{pages}{051915} (\bibinfo{year}{2010}).

\bibitem[{\citenamefont{Malyarchuk et~al.}(2009)\citenamefont{Malyarchuk,
  Castore, and Harrison}}]{Lynn1}
\bibinfo{author}{\bibfnamefont{S.}~\bibnamefont{Malyarchuk}},
  \bibinfo{author}{\bibfnamefont{R.}~\bibnamefont{Castore}}, \bibnamefont{and}
  \bibinfo{author}{\bibfnamefont{L.}~\bibnamefont{Harrison}},
  \bibinfo{journal}{DNA Repair} \textbf{\bibinfo{volume}{8}},
  \bibinfo{pages}{1343} (\bibinfo{year}{2009}).

\bibitem[{\citenamefont{Malyarchuk et~al.}(2008)\citenamefont{Malyarchuk,
  Castore, and Harrison}}]{Lynn2}
\bibinfo{author}{\bibfnamefont{S.}~\bibnamefont{Malyarchuk}},
  \bibinfo{author}{\bibfnamefont{R.}~\bibnamefont{Castore}}, \bibnamefont{and}
  \bibinfo{author}{\bibfnamefont{L.}~\bibnamefont{Harrison}},
  \bibinfo{journal}{Nucleic Acids Res.} \textbf{\bibinfo{volume}{36}},
  \bibinfo{pages}{4872} (\bibinfo{year}{2008}).

\bibitem[{\citenamefont{Sage and Harrison}(2011)}]{Lynn11}
\bibinfo{author}{\bibfnamefont{E.}~\bibnamefont{Sage}} \bibnamefont{and}
  \bibinfo{author}{\bibfnamefont{L.}~\bibnamefont{Harrison}},
  \bibinfo{journal}{Mutat. Res.} \textbf{\bibinfo{volume}{711}},
  \bibinfo{pages}{123} (\bibinfo{year}{2011}).

\bibitem[{\citenamefont{Ward}(1988)}]{Ward1}
\bibinfo{author}{\bibfnamefont{J.}~\bibnamefont{Ward}}, \bibinfo{journal}{Prog.
  Nucleic Acid. Res. Mol. biol.} \textbf{\bibinfo{volume}{35}},
  \bibinfo{pages}{95} (\bibinfo{year}{1988}).

\bibitem[{\citenamefont{Ward}(1995)}]{Ward2}
\bibinfo{author}{\bibfnamefont{J.}~\bibnamefont{Ward}},
  \bibinfo{journal}{Radiat. Res.} \textbf{\bibinfo{volume}{142}},
  \bibinfo{pages}{362} (\bibinfo{year}{1995}).

\bibitem[{\citenamefont{Depken and Schiessel}(2009)}]{nucleosome}
\bibinfo{author}{\bibfnamefont{M.}~\bibnamefont{Depken}} \bibnamefont{and}
  \bibinfo{author}{\bibfnamefont{H.}~\bibnamefont{Schiessel}},
  \bibinfo{journal}{Biophys. J.} \textbf{\bibinfo{volume}{96}},
  \bibinfo{pages}{777} (\bibinfo{year}{2009}).

\bibitem[{\citenamefont{Jakob et~al.}(2003)\citenamefont{Jakob, Scholz, and
  Taucher-Scholz}}]{Jakob}
\bibinfo{author}{\bibfnamefont{B.}~\bibnamefont{Jakob}},
  \bibinfo{author}{\bibfnamefont{M.}~\bibnamefont{Scholz}}, \bibnamefont{and}
  \bibinfo{author}{\bibfnamefont{G.}~\bibnamefont{Taucher-Scholz}},
  \bibinfo{journal}{Radiat. Res.} \textbf{\bibinfo{volume}{159}},
  \bibinfo{pages}{676} (\bibinfo{year}{2003}).

\bibitem[{\citenamefont{Chandrasekhar}(1943)}]{Chandra}
\bibinfo{author}{\bibfnamefont{S.}~\bibnamefont{Chandrasekhar}},
  \bibinfo{journal}{Rev. Mod. Phys.} \textbf{\bibinfo{volume}{15}},
  \bibinfo{pages}{1} (\bibinfo{year}{1943}).

\bibitem[{\citenamefont{Plante and Cucinotta}(2010)}]{Plante}
\bibinfo{author}{\bibfnamefont{I.}~\bibnamefont{Plante}} \bibnamefont{and}
  \bibinfo{author}{\bibfnamefont{F.}~\bibnamefont{Cucinotta}},
  \bibinfo{journal}{Radiat. Environ. Biophys.} \textbf{\bibinfo{volume}{49}},
  \bibinfo{pages}{5} (\bibinfo{year}{2010}).

\bibitem[{\citenamefont{Surdutovich and Solov'yov}(2011)}]{prauger}
\bibinfo{author}{\bibfnamefont{E.}~\bibnamefont{Surdutovich}} \bibnamefont{and}
  \bibinfo{author}{\bibfnamefont{A.}~\bibnamefont{Solov'yov}},
  \bibinfo{journal}{Phys. Rev. Lett.}, \bibinfo{pages}{in preparation}
  (\bibinfo{year}{2011}).

\bibitem[{\citenamefont{Abril et~al.}(2011)\citenamefont{Abril, Garcia-Molina,
  Denton, Kyriakou, and Emfietzoglou}}]{Molina}
\bibinfo{author}{\bibfnamefont{I.}~\bibnamefont{Abril}},
  \bibinfo{author}{\bibfnamefont{R.}~\bibnamefont{Garcia-Molina}},
  \bibinfo{author}{\bibfnamefont{C.}~\bibnamefont{Denton}},
  \bibinfo{author}{\bibfnamefont{I.}~\bibnamefont{Kyriakou}}, \bibnamefont{and}
  \bibinfo{author}{\bibfnamefont{D.}~\bibnamefont{Emfietzoglou}},
  \bibinfo{journal}{Radiat. Res.} \textbf{\bibinfo{volume}{175}},
  \bibinfo{pages}{247} (\bibinfo{year}{2011}).

\bibitem[{\citenamefont{Bug et~al.}(2010)\citenamefont{Bug, Gargioni, Guatelli,
  Incerti, Rabus, Schulte, and Rosenfeld}}]{Marion}
\bibinfo{author}{\bibfnamefont{M.}~\bibnamefont{Bug}},
  \bibinfo{author}{\bibfnamefont{E.}~\bibnamefont{Gargioni}},
  \bibinfo{author}{\bibfnamefont{S.}~\bibnamefont{Guatelli}},
  \bibinfo{author}{\bibfnamefont{S.}~\bibnamefont{Incerti}},
  \bibinfo{author}{\bibfnamefont{H.}~\bibnamefont{Rabus}},
  \bibinfo{author}{\bibfnamefont{R.}~\bibnamefont{Schulte}}, \bibnamefont{and}
  \bibinfo{author}{\bibfnamefont{A.}~\bibnamefont{Rosenfeld}},
  \bibinfo{journal}{Eur. Phys. J. D} \textbf{\bibinfo{volume}{60}},
  \bibinfo{pages}{85} (\bibinfo{year}{2010}).

\bibitem[{\citenamefont{Fuks et~al.}(2011)\citenamefont{Fuks, Horowitz,
  Horowitz, Oster, Marino, Rainer, A.Rosenfeld, and Datz}}]{Roz1}
\bibinfo{author}{\bibfnamefont{E.}~\bibnamefont{Fuks}},
  \bibinfo{author}{\bibfnamefont{Y.}~\bibnamefont{Horowitz}},
  \bibinfo{author}{\bibfnamefont{A.}~\bibnamefont{Horowitz}},
  \bibinfo{author}{\bibfnamefont{L.}~\bibnamefont{Oster}},
  \bibinfo{author}{\bibfnamefont{S.}~\bibnamefont{Marino}},
  \bibinfo{author}{\bibfnamefont{M.}~\bibnamefont{Rainer}},
  \bibinfo{author}{\bibnamefont{A.Rosenfeld}}, \bibnamefont{and}
  \bibinfo{author}{\bibfnamefont{H.}~\bibnamefont{Datz}},
  \bibinfo{journal}{Rad. Prot. Dosim} \textbf{\bibinfo{volume}{143}},
  \bibinfo{pages}{416} (\bibinfo{year}{2011}).

\bibitem[{\citenamefont{Pisacane et~al.}(2010)\citenamefont{Pisacane, Dolecek,
  Malak, Cucinotta, Zaider, Rosenfeld, Rusek, Sivertz, and Dicello}}]{Roz2}
\bibinfo{author}{\bibfnamefont{V.~L.} \bibnamefont{Pisacane}},
  \bibinfo{author}{\bibfnamefont{Q.~E.} \bibnamefont{Dolecek}},
  \bibinfo{author}{\bibfnamefont{H.}~\bibnamefont{Malak}},
  \bibinfo{author}{\bibfnamefont{F.~A.} \bibnamefont{Cucinotta}},
  \bibinfo{author}{\bibfnamefont{M.}~\bibnamefont{Zaider}},
  \bibinfo{author}{\bibfnamefont{A.~B.} \bibnamefont{Rosenfeld}},
  \bibinfo{author}{\bibfnamefont{A.}~\bibnamefont{Rusek}},
  \bibinfo{author}{\bibfnamefont{M.}~\bibnamefont{Sivertz}}, \bibnamefont{and}
  \bibinfo{author}{\bibfnamefont{J.~F.} \bibnamefont{Dicello}},
  \bibinfo{journal}{Rad. Prot. Dosim} \textbf{\bibinfo{volume}{143}},
  \bibinfo{pages}{398} (\bibinfo{year}{2010}).

\bibitem[{\citenamefont{Tung et~al.}(2007)\citenamefont{Tung, Chao, Hsieh, and
  Chan}}]{Tung}
\bibinfo{author}{\bibfnamefont{C.}~\bibnamefont{Tung}},
  \bibinfo{author}{\bibfnamefont{T.}~\bibnamefont{Chao}},
  \bibinfo{author}{\bibfnamefont{H.}~\bibnamefont{Hsieh}}, \bibnamefont{and}
  \bibinfo{author}{\bibfnamefont{W.}~\bibnamefont{Chan}},
  \bibinfo{journal}{Nucl. Inst. Meth B} \textbf{\bibinfo{volume}{262}},
  \bibinfo{pages}{231} (\bibinfo{year}{2007}).

\bibitem[{\citenamefont{Meesungnoen et~al.}(2002)\citenamefont{Meesungnoen,
  Jay-Gerin, Filali-Mouhim, and Mankhetkorn}}]{Meesungnoen02}
\bibinfo{author}{\bibfnamefont{J.}~\bibnamefont{Meesungnoen}},
  \bibinfo{author}{\bibfnamefont{J.-P.} \bibnamefont{Jay-Gerin}},
  \bibinfo{author}{\bibfnamefont{A.}~\bibnamefont{Filali-Mouhim}},
  \bibnamefont{and}
  \bibinfo{author}{\bibfnamefont{S.}~\bibnamefont{Mankhetkorn}},
  \bibinfo{journal}{Radiat. Res.} \textbf{\bibinfo{volume}{158}},
  \bibinfo{pages}{657} (\bibinfo{year}{2002}).

\bibitem[{\citenamefont{Cassie et~al.}(2010)\citenamefont{Cassie, Wroe, Kooy,
  Depauw, Flanz, Paganetti, and Rosenfeld}}]{dosim1}
\bibinfo{author}{\bibfnamefont{B.}~\bibnamefont{Cassie}},
  \bibinfo{author}{\bibfnamefont{A.}~\bibnamefont{Wroe}},
  \bibinfo{author}{\bibfnamefont{H.}~\bibnamefont{Kooy}},
  \bibinfo{author}{\bibfnamefont{N.}~\bibnamefont{Depauw}},
  \bibinfo{author}{\bibfnamefont{J.}~\bibnamefont{Flanz}},
  \bibinfo{author}{\bibfnamefont{H.}~\bibnamefont{Paganetti}},
  \bibnamefont{and}
  \bibinfo{author}{\bibfnamefont{A.}~\bibnamefont{Rosenfeld}},
  \bibinfo{journal}{Med. Phys.} \textbf{\bibinfo{volume}{37}},
  \bibinfo{pages}{311} (\bibinfo{year}{2010}).

\bibitem[{\citenamefont{Kraemer}(2009)}]{Kraemer1}
\bibinfo{author}{\bibfnamefont{M.}~\bibnamefont{Kraemer}},
  \bibinfo{journal}{Nucl. Instr. Meth. B} \textbf{\bibinfo{volume}{267}},
  \bibinfo{pages}{989–992} (\bibinfo{year}{2009}).

\bibitem[{\citenamefont{Tobias et~al.}(2010)\citenamefont{Tobias, Durante,
  Taucher-Scholz, and Jakob}}]{Jakob1}
\bibinfo{author}{\bibfnamefont{F.}~\bibnamefont{Tobias}},
  \bibinfo{author}{\bibfnamefont{M.}~\bibnamefont{Durante}},
  \bibinfo{author}{\bibfnamefont{G.}~\bibnamefont{Taucher-Scholz}},
  \bibnamefont{and} \bibinfo{author}{\bibfnamefont{B.}~\bibnamefont{Jakob}},
  \bibinfo{journal}{Mutat. Res.} \textbf{\bibinfo{volume}{704}},
  \bibinfo{pages}{54} (\bibinfo{year}{2010}).

\end{thebibliography}

\end{document}